# Sentiment Analysis Using Collaborated Opinion Mining

Deepali Virmani, Vikrant Malhotra, Ridhi Tyagi

*Department of Information Technology*
*Bhagwan Parshuram Institute of Technology*
*PSP- 4, Dr. K.N. Katju Marg , Sector 17 Rohini,*
*New Delhi -110089*

[1]deepalivirmani@gmail.com
[2]vikrant.malhotra.1992@gmail.com
[3]ridhi.tyagi.1992@gmail.com

**ABSTRACT**
**Opinion mining and Sentiment analysis have emerged as a field of study since the widespread of World Wide Web and internet. Opinion refers to extraction of those lines or phrase in the raw and huge data which express an opinion. Sentiment analysis on the other hand identifies the polarity of the opinion being extracted. In this paper we propose the sentiment analysis in collaboration with opinion extraction, summarization, and tracking the records of the students. The paper modifies the existing algorithm in order to obtain the collaborated opinion about the students. The resultant opinion is represented as very high, high, moderate, low and very low. The paper is based on a case study where teachers give their remarks about the students and by applying the proposed sentiment analysis algorithm the opinion is extracted and represented.**

## I. INTRODUCTION

There has been a lot of research work that is being done on opinion mining and sentiment analysis. The first question that arises in mind is that what is opinion mining and sentiment analysis? The second thing that one might think would be what is their significance? Opinion is a person's perspective about an issue or an object. Mining refers to extraction of knowledge from raw data or facts [13]. Thus opinion mining is the technique used to extract intelligent information based on a person's opinion from raw data available on internet. Whereas sentiment analysis analyse the polarity of opinion(positive or negative). One might think the need of opinion mining and sentiment analysis as why would one need to know about someone's opinion. The answer lies within the increasing use of internet by people for searching about various products, news, latest information etc. Today people are also placing their comments & opinions on social media so that they can be seen by other people too. Survey has shown that such opinion also affect the people reading those opinions [6,7]. So the remarks related to a product or issue are to be analysed by the associated organization so that they can improve based upon the remarks of people. Opinion mining and sentiment analysis are used to extract such remarks and analyse them on the basis of its polarity respectively. In this paper we try to get collaborated opinion with the help of sentiment analysis. The proposed concept is explained with the help of a case study. The case study is based on the opinion about students extracted from the remarks given by the teachers based on their performance. The remarks are processed with sentimental analysis and opinions (positive or negative) are revealed.

*A. Benefits*

Opinion mining and sentiment analysis not only finds an application in online remark sites but they also can be used as sub-component technology in recommendation systems [1].Opinion mining clubbed with sentiment analysis can be used to produce less of negative feedback and more of positive feedback[1,9]. They can also be used in identifying hostile comments or blogs over the internet by the government. Opinion mining and sentiment analysis also help business analyst to analyse their product remarks and then take effective measures for positive remarks. Another application of opinion mining is in politics. It can be used for identifying people's remark about government candidates standing for elections. It can also be used for analysing people's remark about a government policy.

## II. LITERATURE SURVEY

Many algorithms have been proposed in order to understand and implement opinion mining and sentiment analysis. Researchers have developed models for identifying the polarity of words, sentences and whole document [2].Various tools are also available now for opinion extraction, sentiment analysis and opinion summarization. There have been researches regarding development for better algorithms for such tools.

Kyu, Liang and Chen [2] proposed algorithm for opinion extraction, opinion summarization and tracking the opinion which may be used for multiple languages. The opinion extraction algorithm takes value of opinion holder into consideration whereas in this paper the value of opinion holder is taken to be one.

Ryan [10] proposed document level analysis which has benefit of finer level of classification. He used a structural model with different parameters for document level analysis. He used Viterbi's algorithm for solving the models with different parameters or arguments.

Wilson's [12] approach was to identify contextual polarity for a large subset of sentiment expressions. This approach was a phase-level sentiment analysis.

Hu's and Dave's research focused on extracting opinion from remarks. Hu's [8] research was a product feature based research. It aimed at extracting product features and gave product based summary.

Kin and Hovy [3] in their first model selected a topic and analysed sentiment of remarks using word sentiment classifier with word net. The second model used probability of sentiment words.

### III. PROPOSED CONCEPT

This paper proposes an algorithm for sentiment analysis using collaborated opinion mining which has been explained with the help of a case study. There have been algorithms and models proposed for analysing sentiments and extracting opinions as mentioned in the survey. The basic concept among all the algorithms and models has been identifying sentiment words first. These words are used to find where in the document opinion is present. Then the opinions extracted are analysed to find out the polarity of the opinion. This is the bottom-up approach which is used in most of the algorithms [2]. This paper also implements bottom-up approach to identify opinions present. The proposed algorithm analyses the remarks given by teacher word by word. Analysing the opinion requires analysing of words in a sentence. For such analysis, algorithms proposed have used thesaurus and word net [8, 10]. In this paper, a database of sentiment words has been used for analysis of opinion. Every sentiment word in the database has been given a value. When a sentiment word is detected in a sentence the value saved in the database is used for evaluating the cumulative opinion value.

*A. Proposed Algorithm*

This paper proposes an algorithm for identifying the polarity of remarks. In the existing work polarity of remarks word by word in a sentence was not considered. The proposed work has been explained with help of a case study. A case has been considered wherein a set of teachers give their remarks about a particular student. The algorithm is applied on every remark to identify the polarity of each remark. The algorithm generates a numeric value for the opinion. If the opinion value are high the opinion are considered positive. Lower opinion value represents negative remarks.

The algorithm analyses the remarks word by word [2]. Sentiment words are identified and a combined value is given to each sentence. A database is maintained to identify the sentiment words. The database along with the sentiment word saves an associated value for the opinion word. The value assigned to each sentiment word is based on how much strong or weak sentiment is being used. The value ranges from zero to ten. If a sentiment word emotes strongly positive opinion higher is its value in the database. A sentiment word that represents strong negative opinion lower is its value in the database. When a sentence is analysed, for each sentiment word found in the sentence, its opinion value is fetched from the database. Then the collaborated opinion value of that sentence is estimated. If there is negation in a sentence the value of opinion score is decreased/ increased by a certain amount [2]. The following is the algorithm that is being used.

1. For each word.
2. Check whether it is negation or sentiment word.
3. Every sentiment word is given a value (in the database). A value less than 5 represents negative opinion (e.g. bad). A value greater than 5 represents positive opinion. How much low or how much high value may be decided using thesaurus.
4. If negation is present before sentiment is increased or decreased by 2 depending upon whether the sentiment value is high or low respectively.
5. Now an average is calculated for all the opinion scores calculated for each remark given by teachers. The range of value is 0 to 10. The collaborated opinion score is evaluated as shown:
   If value less than 2
       Very low
   If value greater than 2 but less than 4.5
       Low
   If value greater than 4.5 but less than 5.5
       Moderate
   If value greater than 5.5 but less than 8
       High
   If value greater than 8
       Very high

### IV. CASE STUDY

The following are some of the cases considered for calculating and representing opinion value based on remark given by teacher about a student. Teachers give their remark in the space provided and results are shown.

*Case 1*

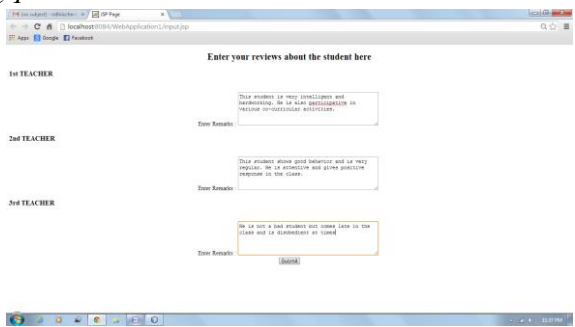

Fig. 1

Remark 1

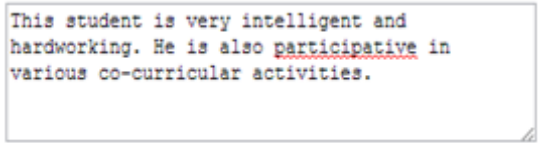

Fig. 1.1

Remark2

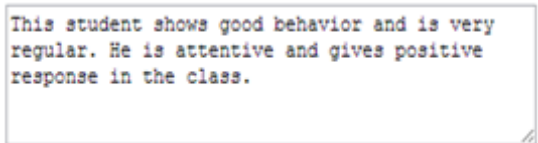

Fig. 1.2

Remark3

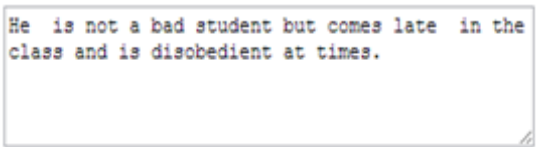

Fig. 1.3

*Result for case 1*

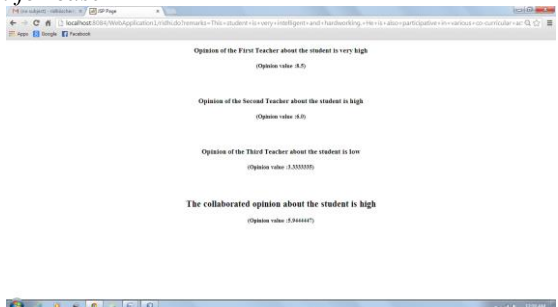

Fig. 2

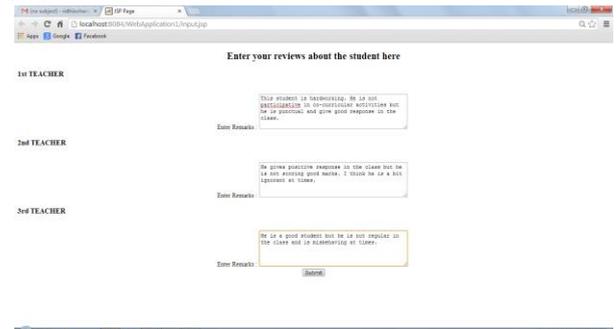

Fig. 2.1

The calculated opinion value comes out to be high since remarks individual response of two teachers comes out to be very high and high. Though the remark of last teacher is low therefore the collaborated opinion is high and not very high.

*Case 2*

Fig. 3

Remark 1

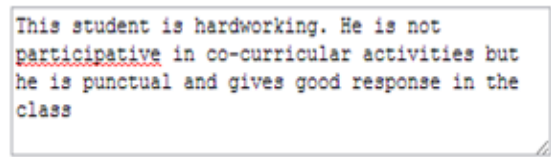

Fig. 3.1

Remark 2

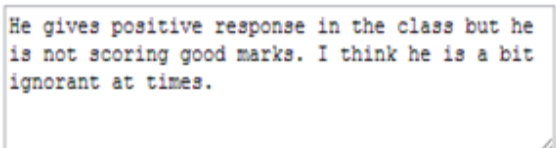

Fig. 3.2

Remark 3

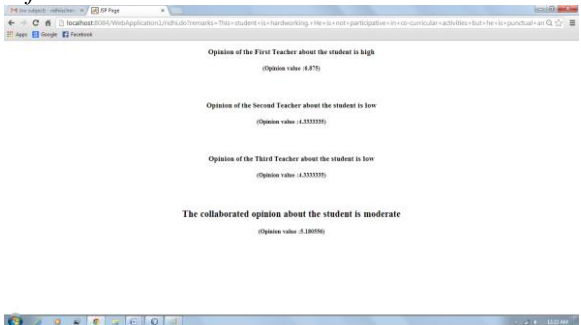

Fig. 3.3

*Result for case 2*

Fig. 4

Opinion of the First Teacher about the student is high

(Opinion value :6.875)

Opinion of the Second Teacher about the student is low

(Opinion value :4.3333335)

Opinion of the Third Teacher about the student is low

(Opinion value :4.3333335)

The collaborated opinion about the student is moderate

(Opinion value :5.180556)

Fig. 4.1

The calculated opinion value comes out to be moderate since two of the teachers gave negative opinions about the student but one gave positive opinion.

*Case 3*

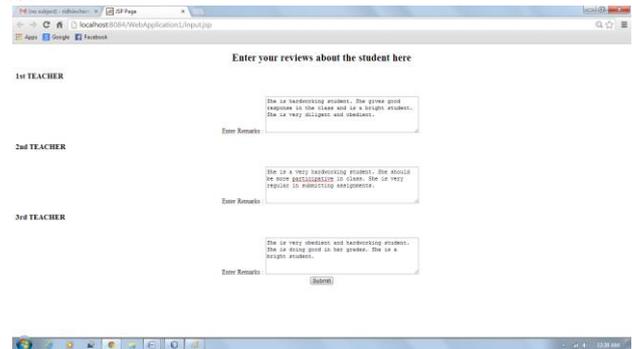

Fig. 5

Remark 1

She is hardworking student. She gives good response in the class and is a bright student. She is very diligent and obedient.

Fig. 5.1

Remark 2

She is a very hardworking student. She should be more participative in class. She is very regular in submitting assignments.

Fig. 5.2

Remark 3

She is very obedient and hardworking student. She is doing good in her grades. She is a bright student.

Fig. 5.3

*Result for case 3*

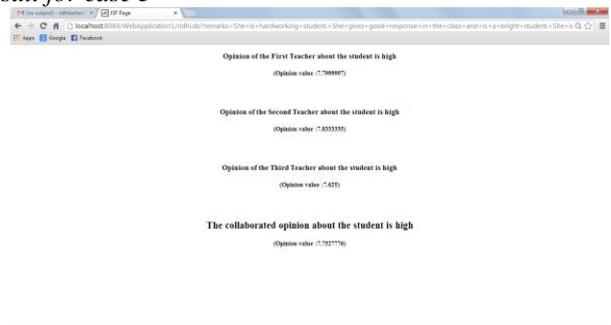

Fig. 6

Opinion of the First Teacher about the student is high

(Opinion value :7.7999997)

Opinion of the Second Teacher about the student is high

(Opinion value :7.8333335)

Opinion of the Third Teacher about the student is high

(Opinion value :7.625)

The collaborated opinion about the student is high

(Opinion value :7.7527776)

Fig. 6.1

The resultant opinion value in this case is high since all the teachers gave positive remarks.

## V. CONCLUSION

This paper proposed algorithm for calculating collaborated opinion value. Various case studies are considered wherein teachers give remark about a student and an average opinion value is calculated. The algorithm compares each word with sentiment and negation in the database. The algorithm is implemented on the basis of score assigned to each sentiment word in the database. The collaborated opinion is evaluated by alalysing teacher's remarks word by word and then implementing the algorithm proposed. The evaluated opinion value for a student can be utilized while giving marks to the student. Reccomendation may be given to a student according to the collaborated opinion value.